\documentclass[aps,prd,twocolumn,10pt,floatfix,nofootinbib,superscriptaddress,showkeys]{revtex4-2}

\usepackage{booktabs}
\usepackage{placeins}
\usepackage{graphicx}
\usepackage{epstopdf}
\usepackage{amssymb}
\usepackage{amsmath}

\usepackage{slashed}
\usepackage{xcolor}
\usepackage[colorlinks=true,linkcolor=blue,citecolor=blue,urlcolor=blue]{hyperref}

\hypersetup{pdftitle={Hadronic production of excited Bc mesons at the LHC},pdfauthor={Hong-Tai Li, Xu-Chang Zheng, Xing-Gang Wu, Jiang Yan}}

\begin{document}

\title{\texorpdfstring{Hadronic production of excited $B_c$ mesons at the LHC}{Hadronic production of excited Bc mesons at the LHC}}

\author{Hong-Tai Li}
\email{liht@cqu.edu.cn}
\author{Xu-Chang Zheng}
\email{zhengxc@cqu.edu.cn}
\author{Xing-Gang Wu}
\email{wuxg@cqu.edu.cn}

\affiliation{Department of Physics, Chongqing Key Laboratory for Strongly Coupled Physics, Chongqing University, Chongqing 401331, China}

\author{Jiang Yan}
\email{yjiang@cqu.edu.cn}
\affiliation{Institute of Theoretical Physics, Chinese Academy of Sciences, Beijing 100049, China}


\begin{abstract}

We study the hadronic production of ground-state and excited $B_c$ mesons at the LHC for $\sqrt{S}=13$ TeV within the framework of NRQCD factorization at leading order in QCD. The calculation includes the $1S$, $1P$, and $2S$ multiplets, with both color-singlet and color-octet contributions to direct $1P$ production. For $p_T>0.2$ GeV, the summed production cross section is $336.04^{+275.80}_{-175.43}$ nb; the quoted uncertainty combines the specified heavy-quark mass and common-scale variations and does not include PDF-set differences. The NNPDF3.0 and MSHT20LO predictions are approximately $23\%$--$24\%$ below the CT18LO result. With the adopted radiative-transition branching fractions, $2S$ feed-down increases the total $1P$ yield from $34.73$ to $75.34$ nb. The relative enhancement differs markedly among the four spectroscopic states, ranging from $38\%$ for $1^1P_1$ to $262\%$ for $1^3P_0$, and it reorders their integrated production hierarchy. Within the set of states included here, the ratio of the $1P$ yield to the inclusive ground-state yield increases from $0.103$ to $0.224$. We present the state-resolved distributions and the dependence of direct production on kinematic cuts, parton distributions, quark masses, and scale choices. These results illustrate the potential impact of $2S$ feed-down on the relative yields of the lowest $P$-wave $B_c$ states.

\end{abstract}


\maketitle

\section{Introduction}

The $B_{c}$ meson is a bound state of a charm quark and a bottom antiquark. Its ground state was first observed at the Tevatron in 1998~\cite{CDF:1998ihx, CDF:1998axz}, and subsequent measurements established its production and decay properties~\cite{D0:2008bqs, D0:2008thm, CDF:2012ksy, LHCb:2012ihf, LHCb:2013xlg, LHCb:2014mvo}. An excited $2S$ structure was reported by ATLAS in 2014~\cite{ATLAS:2014lga}, followed by observations from CMS and LHCb in 2019~\cite{CMS:2019uhm, LHCb:2019bem}. More recently, LHCb observed a structure in the $B_c^+\gamma$ spectrum consistent with orbitally excited $1P$ states~\cite{LHCb:2025uce}. This observation does not resolve the individual spectroscopic states and motivates improved predictions for their production rates. The unequal heavy-quark masses make the $c\bar b$ system a useful probe of heavy-quark dynamics, with extensive theoretical studies of its production and decays~\cite{Chang:1992jb, Chang:1994aw, Kolodziej:1995nv, Berezhnoy:1994ba, Chang:1996jt, Berezhnoy:1996ks, Baranov:1997wy, Baranov:1997sg, Cheung:1999ir, Chang:2003cr, Chang:2004bh, Chang:2005bf, Chang:2005wd, Bi:2016vbt, Chen:2018obq, Berezhnoy:2019yei, Zheng:2019egj, Chang:1992pt, He:2017had}. In this paper, we use $\sqrt{S}=13$ TeV to match the energy of the existing LHCb production measurement~\cite{LHCb:2025study}. Larger data samples at the LHC and its high-luminosity upgrade will extend the opportunities for excited-state spectroscopy~\cite{Campana:2016cqm, ATLAS:2025eii}.

NRQCD factorization separates the perturbative production of a heavy quark-antiquark pair from its nonperturbative evolution into a meson~\cite{Bodwin:1994jh}. For the gluon-gluon fusion channel, the leading-order production rate is of order $\alpha_s^4$. We focus on $g+g\to(c\bar b)[n]+b+\bar c$, including the color-singlet $S$- and $P$-wave configurations and the color-octet $S$-wave configurations that contribute to the physical $P$-wave states~\cite{Chang:2004bh, Chang:2005bf}. The color-octet components are intermediate pair configurations rather than additional asymptotic mesons. Their long-distance matrix elements (LDMEs) introduce nonperturbative uncertainties beyond the heavy-quark mass and scale variations considered below.

The four lowest $P$-wave levels, $B_c(1^1P_1)$ and $B_c(1^3P_J)$ with $J=0,1,2$, feed the $1S$ levels through E1 transitions, $1^1P_1\to1^1S_0+\gamma$ and $1^3P_J\to1^3S_1+\gamma$, the $1^3S_1$ subsequently decaying radiatively to $1^1S_0$~\cite{Eichten:1994gt}. The $2S$ levels decay both radiatively and hadronically into the $1S$ levels, with part of their E1 strength reaching the $1P$ levels; only these radiative branches enter the feed-down model below. We distinguish direct production from the yield including feed-down and count each initially produced meson once in the inclusive ground-state yield.

The BCVEGPY generator implements the amplitudes and phase-space integration needed for $B_c$ hadroproduction~\cite{Chang:2003cq, Chang:2005hq, Wang:2012ah, Chang:2015qea}. Earlier studies established the color-singlet and color-octet contributions to $P$-wave production~\cite{Chang:2004bh, Chang:2005bf}. Eichten and Quigg subsequently combined spectroscopy and decay calculations with BCVEGPY predictions at LHC energies, including direct production and $2S$ feed-down in the radiative signals of the lowest $P$-wave levels~\cite{Eichten:2019}. The same feed-down mechanism enters the production model used in the LHCb analysis~\cite{LHCb:2025study}.

We focus here on the relative production of the four lowest $P$-wave spectroscopic states within a common LO NRQCD setup. For each state, we compare its direct production with its prompt yield including radiative feed-down from the $2S$ states, both in the integrated rate and in the $p_T$, $y$, and $z$ distributions. This comparison shows how feed-down modifies the relative populations of the four states across phase space. We also examine the dependence of the rates, the $1P$ production fractions, and the integrated hierarchy on the PDFs, the heavy-quark masses, the scale choices, and the production-mass prescription. These variations identify which features of the predicted state composition are robust against changes in the production inputs.

The paper is organized as follows. Section~II presents the factorization framework and projection formalism. Section~III gives the numerical inputs, production rates, kinematic distributions, feed-down contributions, and uncertainty estimates. Section~IV summarizes the results and their limitations.

\section{Theoretical framework}

In collinear factorization, the cross section for producing a meson $H$ in proton-proton collisions is given by
\begin{align}
 \sigma(pp\to H+X)
 &=\sum_{i,j}\int_0^1 dx_1\int_0^1 dx_2\, \nonumber\\
 &\quad\times f_{i/p}(x_1,\mu_F)\,f_{j/p}(x_2,\mu_F)\nonumber\\
 &\quad\times \hat\sigma_{ij\to H+X}(x_1,x_2;\mu_R,\mu_F).
\end{align}
Here $f_{i/p}(x,\mu_F)$ is the PDF for parton $i$ carrying momentum fraction $x$ of the proton, $\hat\sigma_{ij\to H+X}$ is the partonic cross section, $\mu_F$ is the factorization scale, and $\mu_R$ is the renormalization scale. In the numerical calculation only the gluon-initiated channel $i=j=g$ is retained, and the two scales are varied together, $\mu_F=\mu_R=\mu$. 

For each meson $H$, the gluon-gluon partonic cross section is further decomposed as
\begin{equation}
 \hat{\sigma}_{gg\to H+X}=\sum_n C_n(\mu_R,\mu_F,\mu_\Lambda)\,\langle\mathcal O^H(n)\rangle_{\mu_\Lambda}.
\label{eq:nrqcd}
\end{equation}
Here $n$ labels the $c\bar b$ configuration, $C_n$ is the corresponding short-distance coefficient, and $\langle\mathcal O^H(n)\rangle_{\mu_\Lambda}$ is the LDME. The configuration $n$ fixes the spin, orbital angular momentum, and color representation of the pair, while $C_n$ encodes its perturbative production. The NRQCD scale $\mu_\Lambda$ separates the short-distance coefficients from the LDMEs and is distinct from the factorization scale $\mu_F$. For the color-singlet components, the LDMEs are proportional, up to normalization factors fixed by the operator convention~\cite{Bodwin:1994jh}, to $|R(0)|^2$ for $S$-wave states and to $|R'(0)|^2$ for $P$-wave states, with $R(0)$ and $R'(0)$ denoting the radial wave function at the origin and its derivative. The color-octet LDMEs are specified separately below. The coefficients $C_n$ are obtained by matching the perturbative amplitudes for an on-shell $c\bar b$ pair in full QCD onto NRQCD at the required order in $\alpha_s$.

For the projected on-shell pair process $g(q_1)+g(q_2)\to(c\bar b)[n](p_1)+b(p_2)+\bar c(p_3)$, we write the perturbative differential cross section as
\begin{equation}
 d\hat\sigma_n^{\rm proj}=\frac{\overline{\sum}|M|^2}{2\hat s}\,d\Phi_3.
 \label{eq:pair-cross-section}
\end{equation}
Here $M$ is the projected pair amplitude. Matching this pair-production result, together with the free-pair matrix elements of Eqs.~\eqref{eq:freepair1} and~\eqref{eq:freepair2}, to NRQCD fixes the normalization of $C_n$ in Eq.~\eqref{eq:nrqcd}; the physical meson cross section then follows with the corresponding meson LDME. The invariant $S=(P_1+P_2)^2$ is the squared proton-proton center-of-mass energy, with $P_{1,2}$ the incoming proton momenta. In collinear kinematics with the proton mass neglected, $q_i=x_iP_i$ ($i=1,2$), so the partonic invariant is $\hat s=(q_1+q_2)^2=x_1x_2S$ and $q_1^2=q_2^2=0$. The symbol $\overline{\sum}$ means that the spins and colors in the initial state are averaged over, and the spins and colors in the final state are summed over. The three-body phase space for the final state can be represented as
\begin{displaymath}
	d{\Phi_3}=(2\pi)^4 \delta^{4}\left(q_1 +q_2 - \sum_{f=1}^3 p_{f}\right)\prod_{f=1}^3 \frac{d^3{p_f}}{(2\pi)^3 2p_f^0}.
\end{displaymath}

There are totally 36 Feynman diagrams, and the projected amplitude for diagram $k$ can be written as
\begin{equation}
	M_{k}^{(2S+1)L_{J}}=C_{k,ij}^{ab}{\bar u}({p_2}) {\cal A}^{(2S+1)L_{J}}{v}({p_3})\,.
\end{equation}
The full amplitude is $M=\sum_k M_k$. The factor $C^{ab}_{k,ij}$ contains the color structure of diagram $k$: $a,b$ label the incoming gluons, and $i,j$ label the outgoing $b$ and $\bar c$. The momenta of these outgoing quarks are $p_2$ and $p_3$. For the constituent color indices $r,s$ of the projected $c\bar b$ pair, the normalized color projectors are
\[
 {\cal C}_{1,rs}=\frac{\delta_{rs}}{\sqrt{N_c}},\qquad
 {\cal C}^{c}_{8,rs}=\sqrt{2}\,T^c_{rs},
\]
with $\operatorname{Tr}(T^cT^{c'})=\delta^{cc'}/2$ and $N_c=3$. The adjoint index $c$ of an octet pair is omitted from $C^{ab}_{k,ij}$ and summed over in the squared amplitude. The spin and orbital structure of the projected $c\bar b$ quark-antiquark pair is contained in ${\cal A}^{(2S+1)L_J}$. For the states considered here,
\begin{align}
	{\cal A}^{S=0,L=0} &= {\cal T}^{(0)}\vert_{q=0}, \\
	{\cal A}^{S=1,L=0} &=  \varepsilon_{\beta}^r (p_1)  {\cal T}^{\beta}\vert_{q=0}, \\
	{\cal A}^{S=0,L=1} &= \varepsilon_{\alpha}^l (p_1) \frac{\partial}{\partial q_{\alpha}}{\cal T}^{(0)}\vert_{q=0}, \\
	{\cal A}^{S=1,L=1} &= \varepsilon_{\alpha \beta}^J (p_1) \frac{\partial}{\partial q_{\alpha}}{\cal T}^{\beta}\vert_{q=0} .
\end{align}
The vectors $\varepsilon_{\beta}^r(p_1)$ and $\varepsilon_{\alpha}^l(p_1)$ describe the spin and orbital polarizations, respectively, while $\varepsilon_{\alpha\beta}^J(p_1)$ is the polarization tensor for the spin-triplet $P$-wave state with $J=0,1,2$. For each of the 36 Feynman diagrams, the Dirac kernel has the form
\begin{align}
	{\cal T}_k &=  {\Gamma_{1k}} \cdot {\Pi_{p_1}(q)} \cdot {\Gamma_{2k}},
\end{align}
where $\Gamma_{1k}$ and $\Gamma_{2k}$ contain the propagators and Dirac structures on either side of the spin projector $\Pi_{p_1}(q)$. The constituent charm-quark and bottom-antiquark momenta are expressed in terms of their relative momentum $q$ as
\begin{equation}
	p_{11}=\frac{m_c}{M}p_1+q ~~{\rm and}~~ p_{12}=\frac{m_b}{M}p_1-q.
\end{equation}
Here $M=m_b+m_c$ is the pair mass used in the leading-order relative-velocity expansion. Its relation to the physical meson masses is discussed in Sec.~\ref{sec:inputs}.
The kernels ${\cal T}^{(0)}$ and ${\cal T}^{\beta}$ use the spin-singlet projector $\Pi^0$ and spin-triplet projector $\Pi^\beta$, respectively. At the order retained here these projectors are
\begin{gather}
	\Pi^0_{p_1}(q)= \frac{-\sqrt{M}}{{4{m_b}{m_c}}}(\slashed{p}_{12}- m_b) \gamma_5 (\slashed{p}_{11} + m_c),\\
	\Pi^{\beta}_{p_1}(q)= \frac{-\sqrt{M}}{{4{m_b}{m_c}}}(\slashed{p}_{12}- m_b) \gamma^{\beta} (\slashed{p}_{11} + m_c).
\end{gather}
The final-state polarization sums can be expressed in terms of the tensor
\begin{equation}
	\Pi_{\alpha \beta}=-g_{\alpha \beta}+\frac{p_{1\alpha}{p_{1\beta}}}{M^2},
\end{equation}
so that the polarization sum for a spin-triplet $S$-wave state or a spin-singlet $P$-wave state is
\begin{equation}
	\sum\limits_{J_z}\varepsilon_{\alpha} \varepsilon^*_{\alpha'}=\Pi_{\alpha \alpha'},
\end{equation}
where $J_z$ can be $r_z$ or $l_z$ for spin and orbit projection respectively. For the three spin-triplet states $B_c(1^3P_J)$ with $J=0,1,2$, the polarization sums are
{\setlength{\arraycolsep}{1pt}
\begin{align}\label{3pja}
	\varepsilon^{(0)}_{\alpha\beta} \varepsilon^{(0)*}_{\alpha'\beta'} &= \frac{1}{3} \Pi_{\alpha\beta}\Pi_{\alpha'\beta'}, \\
	\sum_{J_z}\varepsilon^{(1)}_{\alpha\beta} \varepsilon^{(1)*}_{\alpha'\beta'} &= \frac{1}{2}
	(\Pi_{\alpha\alpha'}\Pi_{\beta\beta'}- \Pi_{\alpha\beta'}\Pi_{\alpha'\beta}), \label{3pjb}\\
	\sum_{J_z}\varepsilon^{(2)}_{\alpha\beta} \varepsilon^{(2)*}_{\alpha'\beta'} &= \frac{1}{2}
	(\Pi_{\alpha\alpha'}\Pi_{\beta\beta'}+ \Pi_{\alpha\beta'}\Pi_{\alpha'\beta})\nonumber \\
	&\quad -\frac{1}{3}
	\Pi_{\alpha\beta}\Pi_{\alpha'\beta'} . \label{3pjc}
\end{align}
}
In the NRQCD expansion, a physical meson contains color-singlet and color-octet Fock components organized by powers of the relative heavy-quark velocity $v$~\cite{Bodwin:1994jh}. The relevant components for the $P$-wave states are
\begin{align}
 |B_c(1^1P_1)\rangle
 &=\mathcal O(v^0)|(c\bar b)[{}^1P_1^{[1]}]\rangle\nonumber\\
 &\quad+\mathcal O(v)|(c\bar b)[{}^1S_0^{[8]}]g\rangle+\cdots,\nonumber\\
 |B_c(1^3P_J)\rangle
 &=\mathcal O(v^0)|(c\bar b)[{}^3P_J^{[1]}]\rangle\nonumber\\
 &\quad+\mathcal O(v)|(c\bar b)[{}^3S_1^{[8]}]g\rangle+\cdots.
 \label{eq:2}
\end{align}
The ellipses denote other Fock components not included in the present calculation. The production cross sections are evaluated in the spectroscopic basis $n^{2S+1}L_J$.

For $H=B_c(n{}^1P_1)$, we introduce the shorthand notation $O_1^H=\langle\mathcal O^H\big({}^1P_1^{[1]}\big)\rangle$ and $O_8^H(\mu_\Lambda)=\langle\mathcal O^H\big({}^1S_0^{[8]}\big)\rangle_{\mu_\Lambda}$. In the spin-summed production convention of Ref.~\cite{Bodwin:1994jh}, the color-singlet LDMEs of the $P$-wave states at leading order in $v$ are
\begin{align*}
   O_1^{B_c(n^1P_1)}  &=\frac{9N_c}{2\pi}|R'_{B_c(nP)}(0)|^2,\\
 \langle\mathcal O^{B_c(n^3P_J)}({}^3P_J^{[1]})\rangle
 &=\frac{3N_c(2J+1)}{2\pi}|R'_{B_c(nP)}(0)|^2.
\end{align*}
These relations receive relative corrections of order $v^2$. In the second line the factor $2J+1$ counts the $J_z$ states of the triplet; in the first line the same factor equals three because the spin-singlet $P$-wave level carries $J=1$. For the lowest $P$-wave multiplet, $|R'_{B_c(1P)}(0)|^2=0.201\,{\rm GeV}^5$ gives $O_1^{B_c(1^1P_1)}=0.864\,{\rm GeV}^5$. The octet matrix element obeys the evolution equation~\cite{Yuan:1994hn}
\begin{equation}
 \mu_\Lambda\frac{dO_8^H(\mu_\Lambda)}{d\mu_\Lambda}
 =\frac{4\alpha_s(\mu_\Lambda)}{27\pi m_{\rm red}^2}O_1^H,
\end{equation}
where $\mu_{\Lambda}$ is the factorization scale in the NRQCD factorization formalism, and the reduced mass is defined as $m_{\mathrm{red}}=m_b m_c/(m_b+m_c)$. Using one-loop running for $\alpha_s$, the leading-logarithmic solution to this RGE is
\begin{align}
 O_8^H(\mu_\Lambda)
 &=O_8^H(\mu_{\Lambda_0})+\frac{8O_1^H}{27\beta_0 m_{\rm red}^2}
 \ln\!\left[\frac{\alpha_s(\mu_{\Lambda_0})}{\alpha_s(\mu_\Lambda)}\right],
 \label{eq.evolsol}
\end{align}
where $\beta_0=11-2n_f/3$ is the first coefficient of the QCD $\beta$-function, $n_f$ denotes the number of active quark flavors, and $\mu_{\Lambda_0}$ is an initial NRQCD factorization scale.

The color-octet LDMEs describe the nonperturbative evolution of an octet pair into a physical $P$-wave meson. The first term on the right-hand side of Eq.~\eqref{eq.evolsol} is the nonperturbative boundary value at the low scale $\mu_{\Lambda0}$, while the second is generated by perturbative evolution. The estimates in Refs.~\cite{Yuan:1994hn,Zheng:2026gluonPwave} are motivated by the limit $\mu_\Lambda\sim m_{\rm red}\gg\mu_{\Lambda_0}$: the evolution term is logarithmically enhanced and can dominate over the boundary contribution, which has no such enhancement. In this approximation, the initial term is neglected and $\alpha_s(\mu_{\Lambda0})\sim1$ is used to estimate the evolution logarithm. For the present numerical analysis, we adopt the phenomenological input
\begin{equation}
 \langle {\cal O}^{B_c(1^1P_1)}(^1S^{[8]}_0)\rangle_{\mu_{\Lambda,\rm ref}}
 =1.4\times10^{-2}\,{\rm GeV}^{3},
 \label{eq:octet-input}
\end{equation}
at the reference scale $\mu_{\Lambda,\rm ref}=m_{\rm red}\simeq1.15$ GeV evaluated with the baseline masses $m_c=1.5$ GeV and $m_b=4.9$ GeV.

According to heavy quark spin symmetry (HQSS)~\cite{Bodwin:1994jh}, the LDMEs satisfy the following relations:
\begin{align}
	\langle {\cal O}^{B_c(n^3P_J)}(^3S^{[8]}_1) \rangle
 &= \frac{2J+1}{3}\langle {\cal O}^{B_c(n^1P_1)}(^1S^{[8]}_0) \rangle.
	\label{eq.HQSS}
\end{align}
These relations hold up to relative corrections of order $v^2$.

With the color projectors specified above, the perturbative matrix elements for an on-shell $c\bar b$ pair, summed over its color and spin states, are at leading order in $\alpha_s$
\begin{align}
	\langle {\cal O}^{(c\bar{b})[^1S^{[8]}_0]}(^1S^{[8]}_0) \rangle &= N_c^2-1, \label{eq:freepair1}\\
	\langle {\cal O}^{(c\bar{b})[^3S^{[8]}_1]}(^3S^{[8]}_1) \rangle &= (d-1)(N_c^2-1), \label{eq:freepair2}
\end{align}
where $N_c=3$ and $d=4$. These free-pair matrix elements fix the color and spin normalization in the perturbative matching for $C_n$; they are distinct from the physical meson LDMEs entering the production cross section.

The direct production rates are evaluated with BCVEGPY~2.2 interfaced to PYTHIA~6.4~\cite{Chang:2015qea}, with the octet normalization specified above. The $2S$ feed-down contributions are obtained by weighting the parent-state rates with the adopted transition branching fractions; the differential prescription is given in Sec.~\ref{sec:feed}.

\begin{table*}[!tp]
\centering
\setlength{\tabcolsep}{6pt}
\caption{Direct production cross sections (nb) for the $1S$, $1P$, and $2S$ $B_c$ states. For the $P$-wave states the color-singlet and color-octet contributions are listed separately; for the $S$-wave states only the color-singlet contribution is included (dashes).}
\label{tb.section}
\begin{tabular}{lrrrrrrrr}
\toprule
Contribution & $1^1S_0$ & $1^3S_1$ & $1^1P_1$ & $1^3P_0$ & $1^3P_1$ & $1^3P_2$ & $2^1S_0$ & $2^3S_1$ \\
\midrule
Color singlet & 52.61 & 131.69 & 6.87 & 2.37 & 5.75 & 14.16 & 33.47 & 83.54 \\
Color octet & - & - & 1.05 & 0.50 & 1.51 & 2.52 & - & - \\
Total & 52.61 & 131.69 & 7.92 & 2.87 & 7.26 & 16.68 & 33.47 & 83.54 \\
\bottomrule
\end{tabular}
\end{table*}

\section{Numerical results and discussion}
\label{results}

\subsection{Input parameters}\label{sec:inputs}

We adopt the following wave-function inputs:
\begin{align}
\vert R_{B_{c}(1S)}(0)\vert^2 &=1.54\,{\rm GeV}^3,\nonumber \\
\vert R'_{B_{c}(1P)}(0)\vert^2 &=0.201\,{\rm GeV}^5,\nonumber \\
\vert R_{B_{c}(2S)}(0)\vert^2 &=0.982\,{\rm GeV}^3.
\end{align}
These values follow the potential-model choices discussed in Refs.~\cite{Chen:1992fq,Chang:1992pt,Chang:1994un,Eichten:1994gt}. The color-octet input and its reference scale are specified in Eq.~\eqref{eq:octet-input} and the accompanying text.

At leading order in the relative-velocity expansion, the on-shell conditions $p_{11}^2=m_c^2$ and $p_{12}^2=m_b^2$, together with $p_1=p_{11}+p_{12}$ at $q=0$, imply $M=m_c+m_b$. We use a unified production-mass prescription, with $m_c=1.5$ GeV, $m_b=4.9$ GeV, and $M=6.4$ GeV for all states. These effective masses enter both the production amplitudes and the phase space; the physical level splittings arise from bound-state dynamics. The state-dependent prescription is used only for the comparison in Table~\ref{smass}, with $(m_c,m_b,M)=(1.275,5.0,6.275)$ GeV for $1S$, $(1.7,5.0,6.7)$ GeV for $1P$, and $(1.633,5.234,6.867)$ GeV for $2S$.

We use CT18LO~\cite{Yan:2022pzl} PDFs with $\alpha_s(m_Z)=0.135$ and set $\mu_F=\mu_R=m_T=\sqrt{M^2+p_T^2}$. The collision energy is $\sqrt{S}=13$ TeV, and the production cut is $p_T>0.2$ GeV, with no additional rapidity restriction. Cuts are applied to the initially produced states. The amplitudes are written for the $c\bar b$ channel, and no additional factor of two for charge-conjugate production is applied to the tabulated rates or event estimates.

Unless otherwise stated, all results use these inputs, the unified mass prescription, and the production cuts specified above. In the input-dependence studies, only the indicated production input is varied; $M=m_c+m_b$ is maintained in the quark-mass scans. The wave-function inputs, the transition branching fractions of Sec.~\ref{sec:feed}, and the color-octet LDME of Eq.~\eqref{eq:octet-input} are held fixed. The LDME reference scale is also kept at its baseline value, $\mu_{\Lambda,\rm ref}\simeq1.15$ GeV, throughout these scans. All plotted distributions show central predictions.

\subsection{Integrated production cross sections}

Table~\ref{tb.section} gives the direct production cross sections for the $1S$, $1P$, and $2S$ multiplets. The singlet and octet rows represent intermediate $c\bar b$ configurations contributing to the same physical final states. In particular, the $^1S_0^{[8]}$ and $^3S_1^{[8]}$ channels contribute to $1^1P_1$ and $1^3P_J$ production, respectively. Although the singlet and octet terms enter $P$-wave production at the same order in NRQCD velocity counting, this counting does not determine their numerical ratio. The singlet and octet contributions summed over the four $1P$ states are $29.15$ and $5.58$ nb: the octet component accounts for approximately $16.1\%$ of direct $1P$ production. State by state it supplies $13.3\%$, $17.4\%$, $20.8\%$, and $15.1\%$ of the $1^1P_1$, $1^3P_0$, $1^3P_1$, and $1^3P_2$ rates, and the three triplet octet rates stand in the ratio $1:3.0:5.0$, close to the $(2J+1)$ weights of Eq.~\eqref{eq.HQSS} once the short-distance coefficient is common to the multiplet. The total for each state is the sum of its displayed singlet and octet contributions. All derived production quantities are calculated from the displayed table entries.

The direct $1S$, $1P$, and $2S$ cross sections sum to approximately $184.30$, $34.73$, and $117.01$ nb, respectively. The direct $P$-wave rates obey
\begin{equation}
 \sigma_{1^1P_1}:\sigma_{1^3P_0}:\sigma_{1^3P_1}:\sigma_{1^3P_2}
 =2.76:1:2.53:5.81.
\end{equation}
Let $\mathcal H$ denote the eight included spectroscopic states and define $\sigma_{\rm sum}=\sum_{H\in\mathcal H}\sigma_H^{\rm direct}$. The baseline sum is $336.04$ nb. All excited states in $\mathcal H$ lie below the open-flavor $BD$ threshold. Their electromagnetic and hadronic transitions to lower $B_c$ levels dominate over competing weak decays; the weak-decay width $\hbar/\tau$ for a lifetime of order $0.5$ ps is about $10^{-12}$ GeV, orders of magnitude below the radiative widths of the corresponding $E1$ transitions. Their inclusive branching fractions to the ground-state $B_c$ are therefore close to unity~\cite{Eichten:1994gt,Eichten:2019}. Accordingly, $\sigma_{\rm sum}$ estimates the inclusive ground-state production cross section from the included states, evaluated with cuts on the initially produced states. We define
\begin{equation}\label{eq:Rdef}
 R_{\rm direct}=\frac{\sum_{H\in1P}\sigma_H^{\rm direct}}{\sigma_{\rm sum}},
 \qquad
 R=\frac{\sum_{H\in1P}\sigma_H^{\rm prompt}}{\sigma_{\rm sum}},
\end{equation}
where $\sigma_H^{\rm prompt}$ is defined in Eq.~\eqref{eq:prompt} and includes direct production plus the specified $2S$ feed-down. The denominator $\sigma_{\rm sum}$ is the sum of the direct rates of the eight included states and already counts each $2S$ parent once; the prompt daughter yields are therefore \emph{not} added to it. Thus $R_{\rm direct}=0.103$.

For an illustrative integrated luminosity of $9\,{\rm fb}^{-1}$ at $13$ TeV, the baseline sum corresponds to $3.02\times10^9$ produced $c\bar b$ mesons (i.e.\ $B_c^+$ in the charge channel specified by the $c\bar b$ amplitude) in the production region defined above. The corresponding reconstructed yield depends on the branching fractions of the selected decay chain and on the acceptance and efficiency for that channel.

\subsection{Direct-production distributions and kinematic-cut dependence}

\begin{figure*}[!tp]
\centering
\includegraphics[trim=75bp 49bp 64bp 42bp,clip,width=\textwidth]{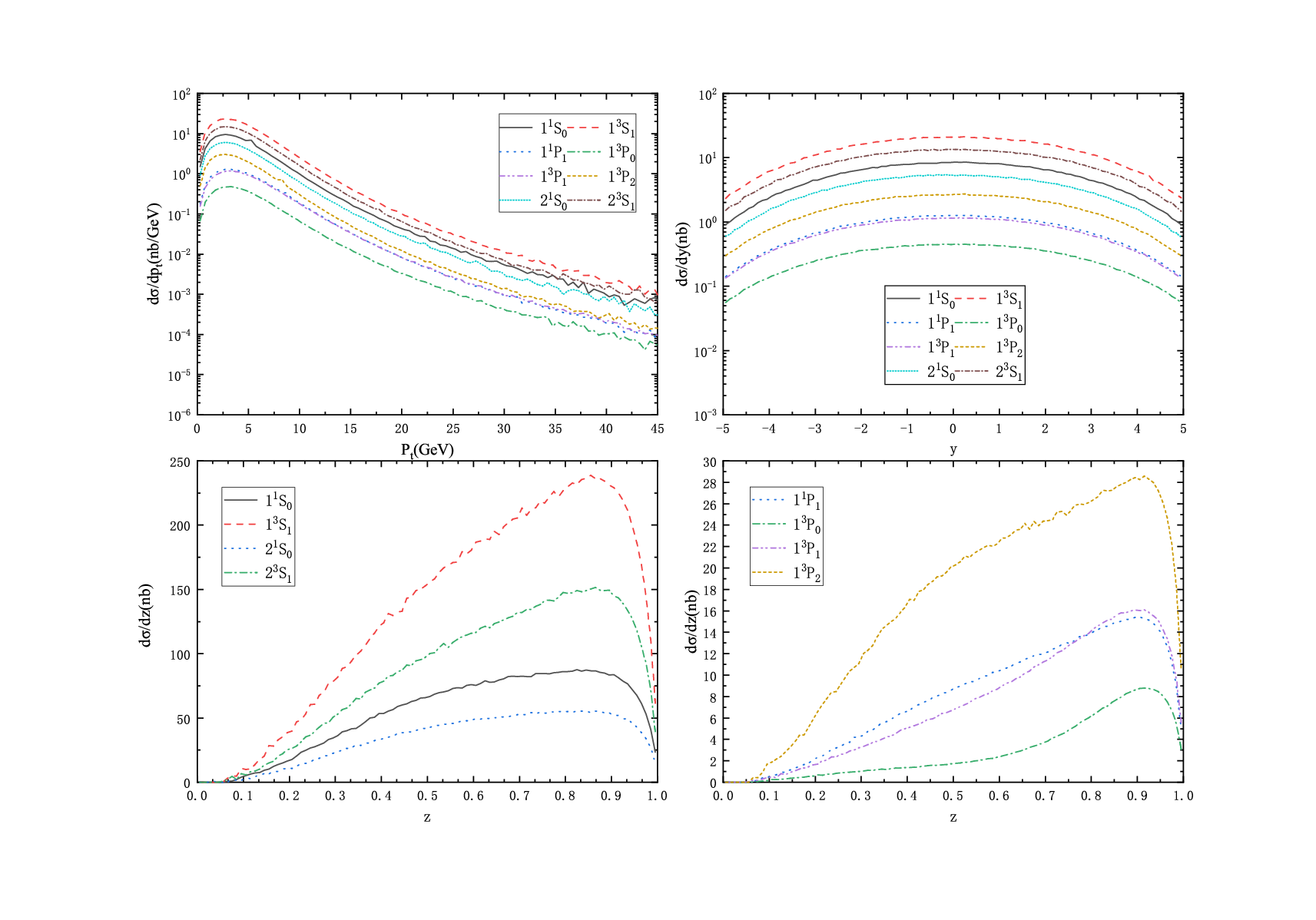}
\caption{Direct-production distributions in transverse momentum $p_T$, rapidity $y$, and energy fraction $z$ for the $1S$, $1P$, and $2S$ $B_c$ states.}  \label{tu1}
\end{figure*}

We study the transverse-momentum, rapidity, and energy-fraction distributions for the included states. Figure~\ref{tu1} shows the differential rates in $p_T$, $y$, and $z$. The rapidity is $y=\frac12\ln[(E+p_z)/(E-p_z)]$ in the proton-proton center-of-mass frame. The energy fraction is defined as $z=2E_{B_c}/\sqrt{\hat s}$, where $E_{B_c}$ is evaluated in the center-of-mass frame of the two incoming partons.

The $p_T$ spectra peak at low transverse momentum and decrease toward larger $p_T$. The rapidity distributions are symmetric about $y=0$, as expected for identical proton beams. The $z$ spectra are shown with separate vertical scales for the larger $S$-wave rates and the smaller $P$-wave rates. Their peaks lie near $z=0.9$, corresponding to $E_{B_c}\simeq0.45\sqrt{\hat s}$ in the partonic center-of-mass frame.

\begin{table}[!htbp]
\centering
\setlength{\tabcolsep}{6pt}
\caption{Direct production cross sections (nb) for $p_T>p_T^{\min}$.}
\label{ptc}
\begin{tabular}{lrrrr}
\toprule
 & \multicolumn{4}{c}{$p_T^{\min}$ (GeV)} \\
\cmidrule(lr){2-5}
State & 2 & 4 & 6 & 8 \\
\midrule
$1^1S_0$ & 42.20 & 24.29 & 11.96 & 5.76 \\
$1^3S_1$ & 106.71 & 62.53 & 30.65 & 14.59 \\
$1^1P_1$ & 6.64 & 4.15 & 2.19 & 1.10 \\
$1^3P_0$ & 2.41 & 1.48 & 0.78 & 0.40 \\
$1^3P_1$ & 6.15 & 3.89 & 2.08 & 1.06 \\
$1^3P_2$ & 13.36 & 7.59 & 3.68 & 1.75 \\
$2^1S_0$ & 26.86 & 15.43 & 7.65 & 3.68 \\
$2^3S_1$ & 67.76 & 39.39 & 19.56 & 9.26 \\
\bottomrule
\end{tabular}
\end{table}

To illustrate the effect of acceptance selections, Tables~\ref{ptc} and \ref{yc} list the state-by-state cross sections for different $p_T$ thresholds and symmetric rapidity intervals. Summing over the eight states, we define

\begin{equation}
 A_T(t)=\frac{\sigma_{\rm sum}(p_T>t)}{\sigma_{\rm sum}(p_T>0.2\,{\rm GeV})}.
\end{equation}
For $t=(2,4,6,8)$ GeV, $A_T=(81\%,47\%,23\%,11\%)$, respectively.

\begin{table}[!htbp]
\centering
\setlength{\tabcolsep}{3pt}
\caption{Direct production cross sections (nb) for $|y|<Y$.}
\label{yc}
\begin{tabular}{lrrrrrr}
\toprule
 & \multicolumn{6}{c}{$Y$} \\
\cmidrule(lr){2-7}
State & 0.5 & 1 & 1.5 & 2 & 3 & 4 \\
\midrule
$1^1S_0$ & 8.19 & 16.09 & 23.93 & 30.93 & 42.05 & 48.50 \\
$1^3S_1$ & 20.57 & 40.74 & 59.76 & 77.33 & 104.27 & 120.71 \\
$1^1P_1$ & 1.23 & 2.45 & 3.61 & 4.64 & 6.28 & 7.28 \\
$1^3P_0$ & 0.44 & 0.87 & 1.28 & 1.65 & 2.26 & 2.64 \\
$1^3P_1$ & 1.14 & 2.23 & 3.27 & 4.23 & 5.73 & 6.68 \\
$1^3P_2$ & 2.58 & 5.20 & 7.58 & 9.86 & 13.33 & 15.50 \\
$2^1S_0$ & 5.23 & 10.32 & 15.14 & 19.61 & 26.71 & 30.85 \\
$2^3S_1$ & 13.09 & 25.61 & 37.68 & 48.71 & 66.23 & 76.94 \\
\bottomrule
\end{tabular}
\end{table}

Similarly, the rapidity acceptance fraction is
\begin{equation}
 A_y(Y)=\frac{\sigma_{\rm sum}(|y|<Y)}{\sigma_{\rm sum}(\text{no rapidity cut})},
\end{equation}
which gives $A_y=(31\%,59\%,79\%,92\%)$ for $Y=(1,2,3,4)$, respectively (Table~\ref{yc}).

Approximately $53\%$ of the baseline yield lies in $0.2<p_T<4$ GeV, and about $79\%$ lies within $|y|<3$.

\begin{table}[!htbp]
\centering
\setlength{\tabcolsep}{16pt}
\caption{Direct production cross sections for $0.2<p_T<20$ GeV and $2<|y|<4.5$.}
\label{sc}
\begin{tabular}{lr}
\toprule
State & $\sigma$ (nb) \\
\midrule
$1^1S_0$ & 19.43 \\
$1^3S_1$ & 48.52 \\
$1^1P_1$ & 2.83 \\
$1^3P_0$ & 1.06 \\
$1^3P_1$ & 2.69 \\
$1^3P_2$ & 6.20 \\
$2^1S_0$ & 12.40 \\
$2^3S_1$ & 31.11 \\
\bottomrule
\end{tabular}
\end{table}

Table~\ref{sc} gives the state-by-state rates in $0.2<p_T<20$ GeV and $2<|y|<4.5$. The sum is $124.24$ nb, or $37.0\%$ of the baseline yield. Direct $1P$ production contributes $12.78$ nb and gives $R_{\rm direct}=0.103$.


\subsection{\texorpdfstring{$2S$ feed-down to the $1P$ states}{2S feed-down to the 1P states}}\label{sec:feed}

The $2S$ multiplet provides a source of $1P$ mesons through radiative transitions. We adopt the following benchmark branching fractions, based on the dominant transitions in Table~XI of Ref.~\cite{Eichten:1994gt}:
\begin{align}
 \mathcal B(2^1S_0\to1^1P_1+\gamma)&=0.09,\nonumber\\
 \mathcal B(2^3S_1\to1^3P_J+\gamma)&=(0.09,0.16,0.20),
 \label{eq:branching}
\end{align}
with $J=(0,1,2)$. For the $J=1$ channels, the levels labeled $2(1^{+\prime})$ and $2(1^+)$ in that reference are assigned approximately to $1^1P_1$ and $1^3P_1$, respectively, according to their dominant spin-preserving transitions. Its principal-quantum-number $2P$ notation corresponds to the lowest, radially labeled $1P$ multiplet used here.

For each included parent $H'$ and daughter $H$, the integrated feed-down estimate is $\sigma_{H'\to H}^{\rm feed}=\sigma_{H'}^{\rm direct}\mathcal B(H'\to H+\gamma)$. In this paper,
\begin{equation}\label{eq:prompt}
 \sigma_H^{\rm prompt}=\sigma_H^{\rm direct}
 +\sum_{H'\in2S}\sigma_{H'\to H}^{\rm feed},\qquad H\in1P.
\end{equation}
The prompt yields presented here combine direct production with the important radiative feed-down contribution from the $2S$ multiplet.

\begin{table}[!htbp]
\centering
\setlength{\tabcolsep}{4pt}
\caption{Direct, $2S$ feed-down, and prompt production cross sections (nb) for the four $1P$ states. Prompt denotes the sum of direct production and $2S$ feed-down.}
\label{2s-p}
\begin{tabular}{lrrrr}
\toprule
Contribution & $1^1P_1$ & $1^3P_0$ & $1^3P_1$ & $1^3P_2$ \\
\midrule
Direct & 7.92 & 2.87 & 7.26 & 16.68 \\
$2S$ feed-down & 3.01 & 7.52 & 13.37 & 16.71 \\
Prompt & 10.93 & 10.39 & 20.63 & 33.39 \\
\bottomrule
\end{tabular}
\end{table}

Table~\ref{2s-p} shows that $2S$ feed-down enhances the $1^1P_1$, $1^3P_0$, $1^3P_1$, and $1^3P_2$ yields by $38\%$, $262\%$, $184\%$, and $100\%$, respectively. Equivalently, feed-down accounts for approximately $28\%$, $72\%$, $65\%$, and $50\%$ of their prompt yields. The strongest relative enhancement occurs for $1^3P_0$, whose direct production is smallest. The integrated hierarchy changes from $1^3P_2>1^1P_1>1^3P_1>1^3P_0$ to $1^3P_2>1^3P_1>1^1P_1>1^3P_0$. The prompt ratios are $1.05:1:1.99:3.21$ in the state order used above. The total feed-down contribution is $40.61$ nb, comparable to and somewhat larger than the direct $1P$ yield of $34.73$ nb. Their sum gives $R=0.224$ as defined in Eq.~\eqref{eq:Rdef}.

For the production cuts in Table~\ref{sc}, the same branching fractions give a summed $2S$ feed-down contribution of $15.12$ nb. Together with the direct $1P$ rate of $12.78$ nb and the total parent rate of $124.24$ nb, this gives $R=0.225$.

LHCb measured $R=0.20\pm0.03\pm0.02\pm0.03$ at $13$ TeV for ground-state $B_c^+$ mesons with $p_T<20$ GeV and $2<y<4.5$~\cite{LHCb:2025study}. The uncertainties are statistical, systematic, and due to the models used to extract the combined $1P$ yield, respectively. LHCb also quotes BCVEGPY predictions of $R=0.17$--$0.19$ that include color-octet production and radiative $2S$ feed-down~\cite{LHCb:2025study}. In our estimate of $R=0.225$, the cuts in Table~\ref{sc} apply to the initially produced mesons. The forward and backward regions contribute equally and their common factor cancels in the ratio.

We next consider the differential distributions. The feed-down contribution is evaluated bin by bin from the $2S$ parent histograms using the transition branching fractions. For $X=p_T$, $y$, or $z$, the prescription is
\begin{align}
 \frac{d\sigma_H^{\rm prompt}}{dX}
 &=\frac{d\sigma_H^{\rm direct}}{dX}\nonumber\\
 &\quad+\sum_{H'\in2S}\mathcal B(H'\to H+\gamma)
 \left.\frac{d\sigma_{H'}^{\rm direct}}{dX'}\right|_{X'=X}.
 \label{eq:feed-spectrum}
\end{align}
Here the daughter is assigned the parent's value of $X$, with the production cuts applied to the parent. This prescription neglects decay recoil and angular correlations. For a parent of mass $M_{H'}$ and daughter of mass $M_H$, the photon momentum in the parent's rest frame is $q_\gamma^*=(M_{H'}^2-M_H^2)/(2M_{H'})$. The masses of Ref.~\cite{Eichten:1994gt} give $q_\gamma^*\simeq0.12$--$0.20$ GeV and $q_\gamma^*/M_{H'}\simeq0.017$--$0.028$ for the included transitions.

\begin{figure}[htbp]
\centering
\includegraphics[trim=55bp 25bp 80bp 45bp,clip,width=0.45\textwidth]{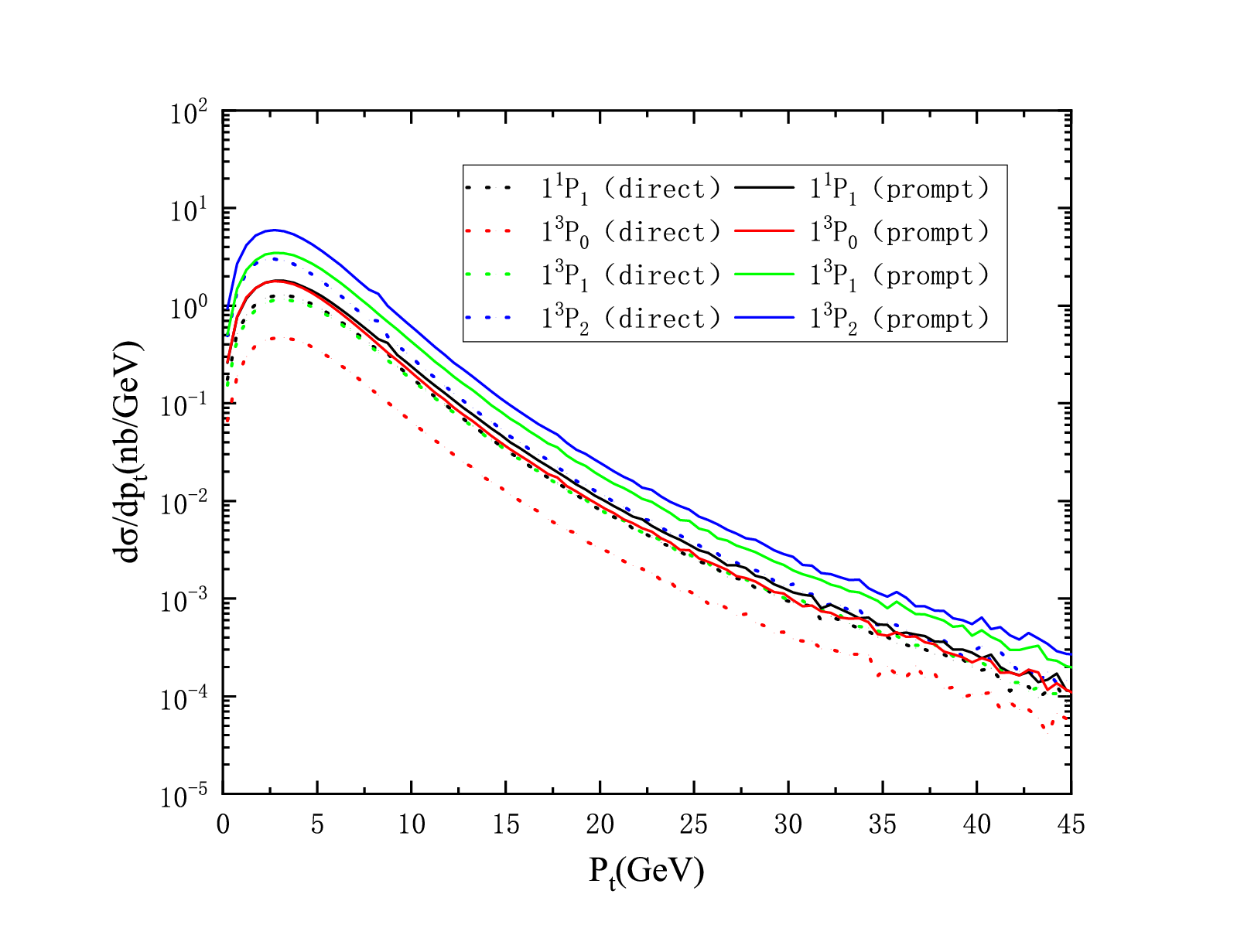}
\caption{Transverse-momentum distributions for the four $1P$ states. Dashed and solid curves show direct and prompt production (direct plus $2S$ feed-down), respectively. Colors distinguish the states.}
\label{P1tu}
\end{figure}

\begin{figure}[htbp]
\centering
\includegraphics[trim=55bp 25bp 80bp 45bp,clip,width=0.45\textwidth]{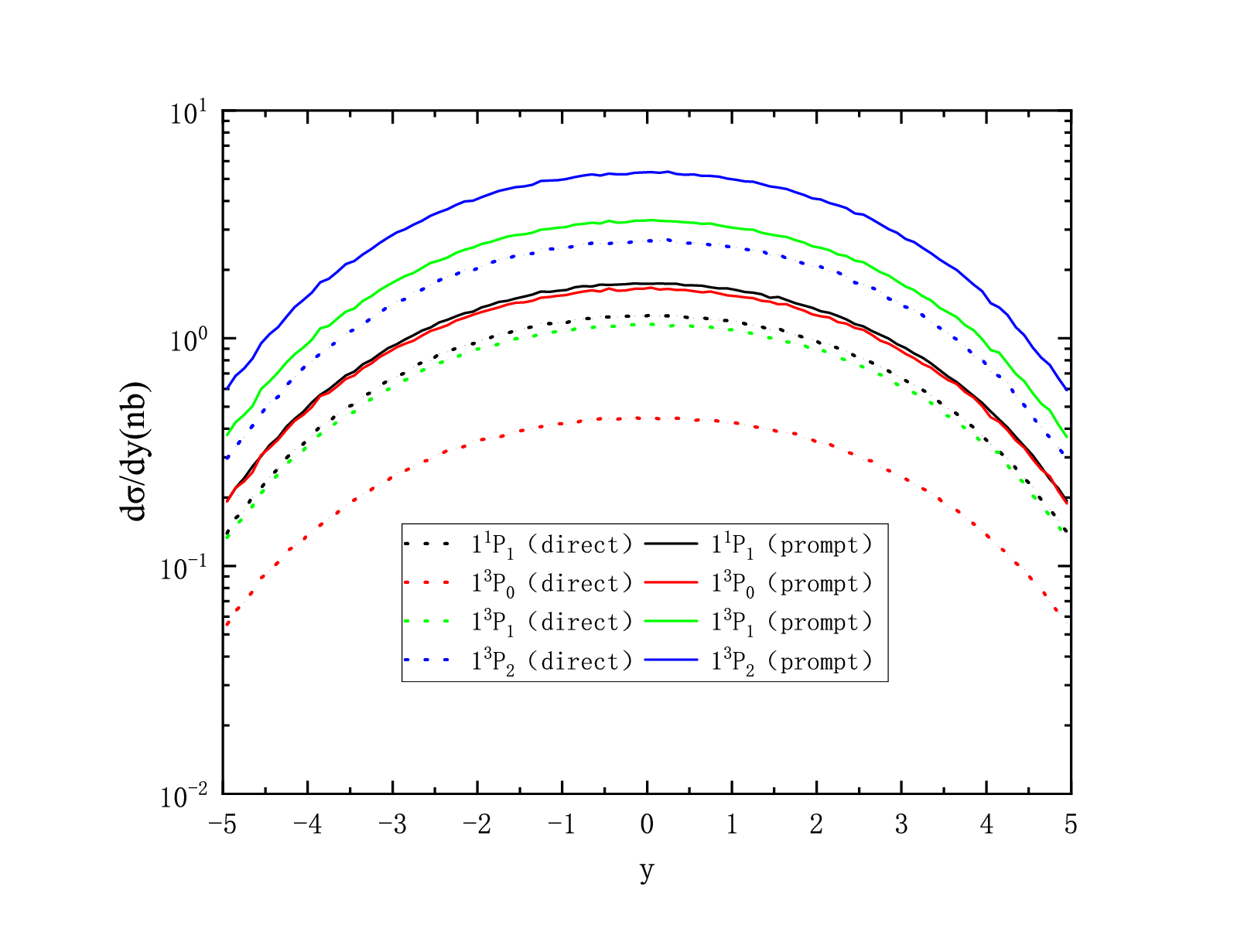}
\caption{Rapidity distributions for the four $1P$ states. Line styles and colors follow Fig.~\ref{P1tu}.}
\label{rap1tu}
\end{figure}

\begin{figure}[htbp]
\centering
\includegraphics[trim=55bp 25bp 80bp 45bp,clip,width=0.45\textwidth]{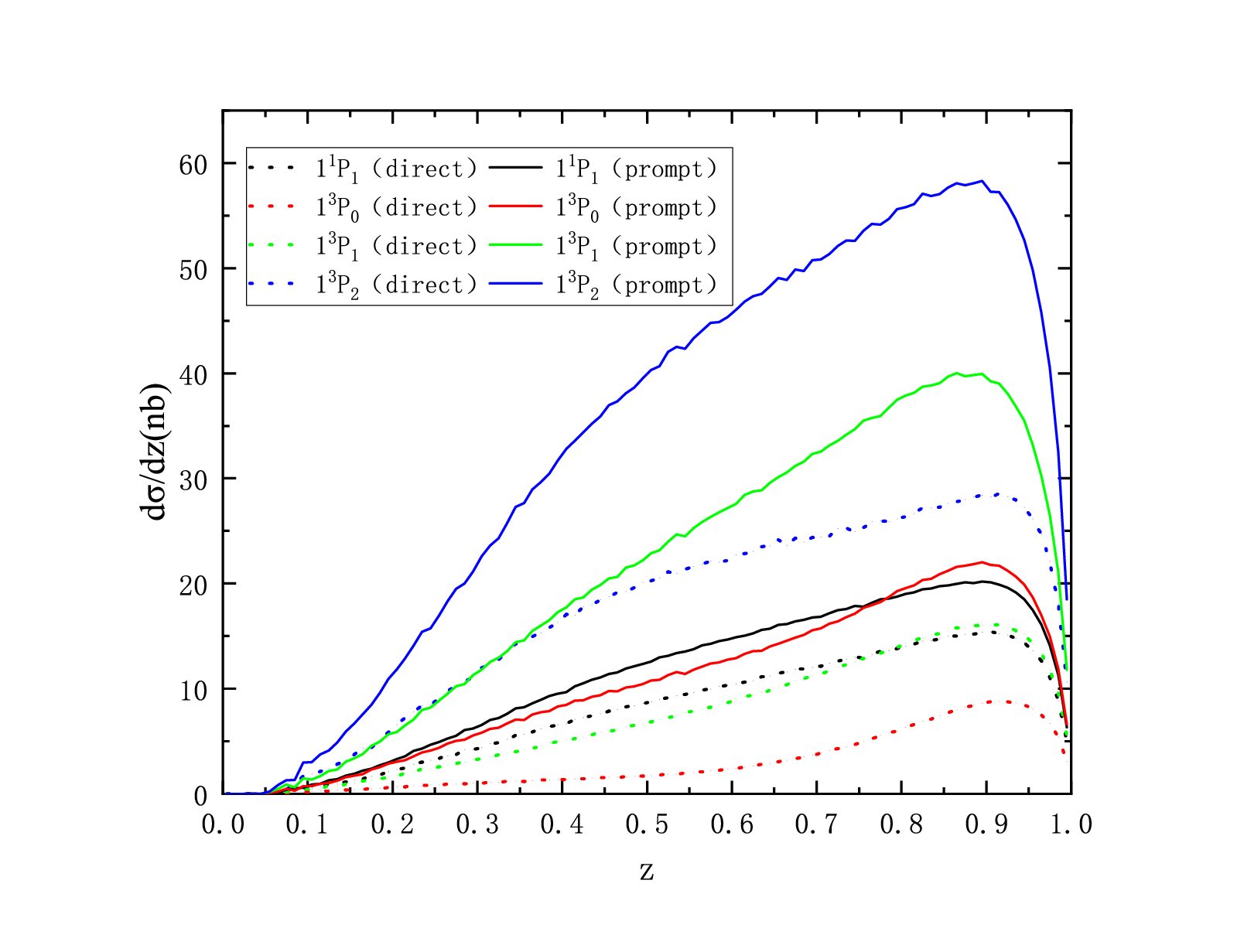}
\caption{Energy-fraction ($z$) distributions for the four $1P$ states. Line styles and colors follow Fig.~\ref{P1tu}.}
\label{z1tu}
\end{figure}

Figures~\ref{P1tu}--\ref{z1tu} compare direct and prompt distributions for the four $1P$ spectroscopic states, using the branching fractions of Eq.~\eqref{eq:branching} and the feed-down prescription of Eq.~\eqref{eq:feed-spectrum}. Each color labels one state; dashed and solid curves show direct production and the direct-plus-$2S$ contribution, respectively. Direct production includes both singlet and octet channels.

The $p_T$ spectra in Fig.~\ref{P1tu} peak at a few GeV and decrease rapidly at larger $p_T$. The prompt curves lie above their direct counterparts, with different relative enhancements for the four states. The integrated ordering discussed above need not hold in every momentum bin.

The rapidity distributions in Fig.~\ref{rap1tu} are broad and symmetric about $y=0$, as expected for identical proton beams. Feed-down increases the normalization throughout the plotted central region.

The $z$ distributions in Fig.~\ref{z1tu} increase toward a maximum near $z=0.9$ and then fall near the endpoint. The prompt $1^1P_1$ and $1^3P_0$ curves cross at large $z$: the latter becomes larger in that region even though its integrated yield is smaller. This illustrates why integrated production fractions alone do not fix the relative contributions in a selected kinematic interval.

\subsection{Input dependence and quoted uncertainties}\label{sec:uncertainty}

We examine the dependence of direct production on the PDF choice, heavy-quark masses, and common scale $\mu_F=\mu_R$. The combined error includes only the stated quark-mass ranges and the common-scale choices $0.5m_T$ and $2m_T$. PDF-set differences and the alternative mass prescription are displayed separately.

Table~\ref{pdf} compares the central predictions of CT18LO~\cite{Yan:2022pzl}, MMHT2014lo~\cite{Harland-Lang:2014zoa}, NNPDF3.0~\cite{NNPDF:2014otw}, and MSHT20LO~\cite{Bailey:2020ooq}, using the associated strong-coupling prescription for each PDF choice.

\begin{table}[!htbp]
\centering
\setlength{\tabcolsep}{3pt}
\caption{Direct production cross sections (nb) for different central LO PDF sets. MMHT14 denotes MMHT2014lo.}
\label{pdf}
\begin{tabular}{lrrrr}
\toprule
State & CT18LO & MMHT14 & NNPDF3.0 & MSHT20 \\
\midrule
$1^1S_0$ & 52.61 & 52.87 & 40.72 & 39.96 \\
$1^3S_1$ & 131.69 & 131.91 & 101.67 & 99.95 \\
$1^1P_1$ & 7.92 & 7.93 & 6.12 & 6.01 \\
$1^3P_0$ & 2.87 & 2.89 & 2.23 & 2.19 \\
$1^3P_1$ & 7.26 & 7.25 & 5.60 & 5.51 \\
$1^3P_2$ & 16.68 & 16.68 & 12.88 & 12.66 \\
$2^1S_0$ & 33.47 & 33.54 & 25.84 & 25.45 \\
$2^3S_1$ & 83.54 & 83.65 & 64.65 & 63.42 \\
\bottomrule
\end{tabular}
\end{table}

The CT18LO and MMHT2014lo predictions agree to within approximately $1\%$. The summed rates for NNPDF3.0 and MSHT20LO are $22.7\%$ and $24.1\%$ lower than CT18LO, respectively. Differences in the gluon PDFs and the associated strong-coupling prescriptions can both contribute to this spread.

\begin{table}[!htbp]
\centering
\setlength{\tabcolsep}{3pt}
\caption{Direct production cross sections (nb) for different $c$-quark masses.}
\label{mass1}
\begin{tabular}{lrrrrr}
\toprule
 & \multicolumn{5}{c}{$m_c$ (GeV)} \\
\cmidrule(lr){2-6}
State & 1.3 & 1.4 & 1.5 & 1.6 & 1.7 \\
\midrule
$1^1S_0$ & 81.26 & 64.56 & 52.61 & 43.10 & 35.87 \\
$1^3S_1$ & 214.76 & 166.50 & 131.69 & 105.19 & 85.94 \\
$1^1P_1$ & 16.02 & 11.10 & 7.92 & 5.78 & 4.31 \\
$1^3P_0$ & 5.38 & 3.88 & 2.87 & 2.18 & 1.68 \\
$1^3P_1$ & 14.37 & 10.08 & 7.26 & 5.35 & 4.05 \\
$1^3P_2$ & 34.08 & 23.55 & 16.68 & 12.18 & 9.10 \\
$2^1S_0$ & 51.75 & 41.25 & 33.47 & 27.43 & 22.78 \\
$2^3S_1$ & 136.09 & 105.29 & 83.54 & 67.00 & 54.34 \\
\bottomrule
\end{tabular}
\end{table}

Tables~\ref{mass1} and \ref{mass2} give the mass dependence over $m_c=1.5\pm0.2$ GeV and $m_b=4.9\pm0.4$ GeV, respectively.

\begin{table}[!htbp]
\centering
\setlength{\tabcolsep}{3pt}
\caption{Direct production cross sections (nb) for different $b$-quark masses.}
\label{mass2}
\begin{tabular}{lrrrrr}
\toprule
 & \multicolumn{5}{c}{$m_b$ (GeV)} \\
\cmidrule(lr){2-6}
State & 4.5 & 4.7 & 4.9 & 5.1 & 5.3 \\
\midrule
$1^1S_0$ & 71.06 & 60.68 & 52.61 & 45.40 & 39.54 \\
$1^3S_1$ & 173.09 & 150.13 & 131.69 & 115.24 & 101.65 \\
$1^1P_1$ & 10.67 & 9.19 & 7.92 & 6.88 & 6.00 \\
$1^3P_0$ & 4.03 & 3.39 & 2.87 & 2.45 & 2.12 \\
$1^3P_1$ & 9.77 & 8.39 & 7.26 & 6.30 & 5.52 \\
$1^3P_2$ & 22.35 & 19.32 & 16.68 & 14.58 & 12.79 \\
$2^1S_0$ & 45.37 & 38.82 & 33.47 & 28.97 & 25.33 \\
$2^3S_1$ & 110.08 & 95.42 & 83.54 & 73.15 & 64.60 \\
\bottomrule
\end{tabular}
\end{table}

The separate shifts in the summed rate from the charm- and bottom-mass scans are
\begin{align}
 \sigma_{\rm sum}&=336.04^{+217.67}_{-117.97}\ {\rm nb}
 &&(m_c\text{ scan}),\label{eq:mcscan}\\
 \sigma_{\rm sum}&=336.04^{+110.38}_{-78.49}\ {\rm nb}
 &&(m_b\text{ scan}).\label{eq:mbscan}
\end{align}

Increasing (decreasing) $m_c$ by $0.2$ GeV reduces (increases) the summed rate by approximately $35\%$ ($65\%$). Increasing (decreasing) $m_b$ by $0.4$ GeV changes it by approximately $-23\%$ ($+33\%$). These variations show that the charm-mass dependence is stronger than the bottom-mass dependence within the specified ranges.

Table~\ref{smass} compares the unified and state-dependent production-mass prescriptions.

		\begin{table}[!htbp]
\centering
\setlength{\tabcolsep}{8pt}
\caption{Direct production cross sections (nb) for the unified and state-dependent production-mass prescriptions.}
\label{smass}
\begin{tabular}{lrr}
\toprule
State & Unified & State dependent \\
\midrule
$1^1S_0$ & 52.61 & 80.21 \\
$1^3S_1$ & 131.69 & 213.70 \\
$1^1P_1$ & 7.92 & 5.21 \\
$1^3P_0$ & 2.87 & 2.06 \\
$1^3P_1$ & 7.26 & 5.29 \\
$1^3P_2$ & 16.68 & 11.01 \\
$2^1S_0$ & 33.47 & 20.33 \\
$2^3S_1$ & 83.54 & 50.62 \\
\bottomrule
\end{tabular}
\end{table}

Relative to the unified prescription, the state-dependent prescription increases the $1^1S_0$ and $1^3S_1$ rates by approximately $52\%$ and $62\%$, respectively. The four $1P$ rates decrease by approximately $27\%$--$34\%$, while the $2S$ rates decrease by approximately $39\%$.

Table~\ref{mu} compares the common scale choices $\mu_F=\mu_R=0.5m_T$, $m_T$, $2m_T$, $\sqrt{\hat s}/2$, and $\sqrt{\hat s}$. Varying the common scale changes both the strong coupling and the PDFs.

\begin{table}[!htbp]
\centering
\setlength{\tabcolsep}{3pt}
\caption{Direct production cross sections (nb) for different common scales $\mu_F=\mu_R$.}
\label{mu}
\begin{tabular}{lrrrrr}
\toprule
State & $0.5m_T$ & $m_T$ & $2m_T$ & $\sqrt{\hat s}/2$ & $\sqrt{\hat s}$ \\
\midrule
$1^1S_0$ & 72.23 & 52.61 & 36.44 & 37.81 & 26.60 \\
$1^3S_1$ & 182.13 & 131.69 & 90.78 & 93.36 & 65.57 \\
$1^1P_1$ & 10.99 & 7.92 & 5.47 & 5.67 & 3.98 \\
$1^3P_0$ & 3.91 & 2.87 & 2.03 & 2.25 & 1.60 \\
$1^3P_1$ & 9.99 & 7.26 & 5.06 & 5.39 & 3.80 \\
$1^3P_2$ & 23.17 & 16.68 & 11.54 & 11.83 & 8.30 \\
$2^1S_0$ & 46.19 & 33.47 & 23.32 & 24.09 & 16.97 \\
$2^3S_1$ & 115.88 & 83.54 & 57.96 & 59.24 & 41.72 \\
\bottomrule
\end{tabular}
\end{table}

For example, the $1^1S_0$ rate decreases from $52.61$ nb at $\mu=m_T$ to $26.60$ nb at $\mu=\sqrt{\hat s}$, a reduction of $49.4\%$. The summed rate changes by approximately $+38.2\%$ at $0.5m_T$ and $-30.8\%$ at $2m_T$.

For each state, we combine the positive and negative shifts from $m_c=1.5\pm0.2$ GeV, $m_b=4.9\pm0.4$ GeV, and $\mu=0.5m_T,2m_T$ separately in quadrature. Thus $\Delta_\pm=\sqrt{(\Delta_{m_c,\pm})^2+(\Delta_{m_b,\pm})^2+(\Delta_{\mu,\pm})^2}$. This prescription gives
\begin{align}
\sigma_{(1^{1}S_{0})} &= \left(52.61^{+39.32}_{-26.69}\right)\;{\rm nb}, \nonumber\\
\sigma_{(1^{3}S_{1})} &= \left(131.69^{+105.64}_{-68.33}\right)\;{\rm nb}, \nonumber\\
\sigma_{(1^{1}P_{1})} &= \left(7.92^{+9.09}_{-4.77}\right)\;{\rm nb}, \nonumber\\
\sigma_{(1^{3}P_{0})} &= \left(2.87^{+2.95}_{-1.64}\right)\;{\rm nb}, \nonumber\\
\sigma_{(1^{3}P_{1})} &= \left(7.26^{+8.02}_{-4.26}\right)\;{\rm nb}, \nonumber\\
\sigma_{(1^{3}P_{2})} &= \left(16.68^{+19.42}_{-9.95}\right)\;{\rm nb}, \nonumber\\
\sigma_{(2^{1}S_{0})} &= \left(33.47^{+25.25}_{-16.84}\right)\;{\rm nb}, \nonumber\\
\sigma_{(2^{3}S_{1})} &= \left(83.54^{+67.17}_{-43.19}\right)\;{\rm nb}.
\end{align}

For the summed yield, the shift from each input variation is first summed across all states, retaining the correlation induced by that common input. The three source shifts are then combined in quadrature, giving
\begin{align}
	\label{eq:sum}\sigma_{\rm sum} &= 336.04^{+275.80}_{-175.43}\;{\rm nb}.
\end{align}
The charm-mass scan gives the largest contribution to this combined error, followed by the common-scale and bottom-mass variations.

One further source lies outside the quadrature of Eq.~\eqref{eq:sum}: the NNPDF3.0 and MSHT20LO predictions in Table~\ref{pdf} are $22.7\%$ and $24.1\%$ below CT18LO, respectively, whereas MMHT2014lo differs by only $+0.2\%$. Finally, the calculation is leading order in $\alpha_s$; the $+38\%$ and $-31\%$ common-scale shifts stand as the conventional proxy for the missing higher orders, and the rates quoted here are LO benchmarks rather than predictions with a controlled error.

The sensitivity of $R_{\rm direct}$ and $R$ can be assessed directly from Tables~\ref{pdf}--\ref{mu}, recalculating both numerator and denominator for each column and using the fixed branching fractions in Eq.~\eqref{eq:branching}. The prompt numerator is the summed direct $1P$ rate plus $0.09\,\sigma_{2^1S_0}^{\rm direct}+0.45\,\sigma_{2^3S_1}^{\rm direct}$. The PDF sets give $R_{\rm direct}=0.1032$--$0.1034$ and $R=0.2240$--$0.2243$; the bottom-mass scan gives $R_{\rm direct}=0.1026$--$0.1049$ and $R=0.2242$--$0.2251$; the common-scale choices give $R_{\rm direct}=0.1034$--$0.1049$ and $R=0.2242$--$0.2254$. Relative to the baseline values, all these variations are below $2\%$ for $R_{\rm direct}$ and $0.6\%$ for $R$.

The charm-mass scan $m_c=1.5\pm0.2$ GeV gives the wider ranges $R_{\rm direct}\simeq0.088$--$0.126$ and $R\simeq0.209$--$0.245$, with the larger ratios at the smaller mass. For the state-dependent prescription, summing the corresponding column of Table~\ref{smass} gives $\sigma_{\rm sum}=388.43$ nb, a direct $1P$ rate of $23.57$ nb, and a prompt $1P$ rate of $48.18$ nb. The corresponding ratios are $R_{\rm direct}\simeq0.061$ and $R\simeq0.124$. This substantial shift reflects the different effective production masses assigned to the multiplets, including $m_c=1.275$ GeV for $1S$ and $1.7$ GeV for $1P$. The state-dependent prescription preserves the baseline prompt ordering, $1^3P_2>1^3P_1>1^1P_1>1^3P_0$, despite changing the overall $1P$ fraction.

At every tabulated variation point, the central prompt $1^3P_2$ yield remains the largest, followed by $1^3P_1$, and both exceed the $1^1P_1$ and $1^3P_0$ yields. The relative ordering of the two smaller yields is less stable. At $m_c=1.6$ GeV they are nearly equal, $\sigma^{\rm prompt}_{1^1P_1}\simeq8.25$ nb and $\sigma^{\rm prompt}_{1^3P_0}\simeq8.21$ nb; at $m_c=1.7$ GeV their central values interchange, becoming $6.36$ and $6.57$ nb, respectively.

\FloatBarrier
\section{Conclusions}\label{summary}

We have calculated the gluon-fusion cross sections for the $1S$, $1P$, and $2S$ $B_c$ multiplets at $\sqrt{S}=13$ TeV to leading order in $\alpha_s$ and in the relative-velocity expansion. For $p_T>0.2$ GeV and the common production-mass prescription, the summed cross section is $336.04^{+275.80}_{-175.43}$ nb, where the uncertainty covers the specified quark-mass and common-scale variations. An integrated luminosity of $9\,{\rm fb}^{-1}$ at this energy would correspond to $(3.02^{+2.48}_{-1.58})\times10^9$ produced mesons before decay branching fractions and experimental efficiencies.

With the adopted branching fractions, radiative $2S$ feed-down adds $40.61$ nb to the direct $1P$ cross section of $34.73$ nb and raises the $1P$ fraction from $0.103$ to $0.224$. Relative to direct production, the $1^1P_1$, $1^3P_0$, $1^3P_1$, and $1^3P_2$ yields increase by $38\%$, $262\%$, $184\%$, and $100\%$, respectively. Feed-down supplies $72\%$ of the prompt $1^3P_0$ yield and brings $1^3P_1$ above $1^1P_1$ at the baseline inputs. The $1^3P_2$ and $1^3P_1$ states have the largest and second-largest prompt yields, respectively, throughout the production-parameter scans. At $m_c=1.7$ GeV, the prompt $1^3P_0$ yield exceeds that of $1^1P_1$, reversing their baseline ordering.

The $1P$ fractions $R_{\rm direct}$ and $R$ vary by less than $2\%$ and $0.6\%$, respectively, under the PDF, common-scale, and bottom-mass variations. The charm-mass scan gives wider ranges, $R_{\rm direct}=0.088$--$0.126$ and $R=0.209$--$0.245$, while the state-dependent mass prescription gives $0.061$ and $0.124$, respectively. With the adopted color-octet LDME, the octet channels account for $13\%$--$21\%$ of the direct production cross section, depending on the $1P$ state. The state-resolved prompt cross sections provide estimates of the relative production weights of the four levels for studies of the $B_c^+\gamma$ mass spectrum.

\begin{acknowledgments} This work was supported in part by the Natural Science Foundation of China under Grants No. 12575080, No. 12675111, No. 12547115 and No. 12547101, by the Fundamental Research Funds for the Central Universities under Grant No. 2026CDJZKPT-10. \end{acknowledgments}


\end{document}